\documentclass[11pt]{article}

\usepackage{tlatex}
\usepackage{amsfonts,amsmath}
\usepackage{url}
\usepackage{xspace}

\newcommand{\ma}[1]{\ensuremath{#1}\xspace}

\def\mdef#1#2{\def#1{\ma{#2}}}
\mdef\a{\alpha}
\mdef{\is}{\colon\!}

\newcommand\newmeta[2]{\def#1##1{\ensuremath{#2_{##1}}}}
\newmeta\p{\phi}

\newcommand{\msc}[1]{\mbox{\textsc{#1}}\xspace}
\newcommand{\tlaplus}{$\msc{TLA}^{\scriptstyle{+}}$\xspace}
\newcommand{\tlaps}{\msc{tlaps}}
\newcommand{\Tlaplus}{{\upshape TLA}$^+$\xspace}
\newcommand{\TLAPS}{\msc{tlaps}}

\newcommand{\SMTLIB}{\msc{smt-lib}}
\newcommand{\SMT}{\msc{smt}}
\newcommand{\ZF}{\msc{zf}}

\newcommand{\FOL}{\msc{fol}}
\newcommand{\MSFOL}{\msc{ms-fol}}

\newcommand{\TPTP}{\msc{tptp}}
\newcommand{\AUFLIA}{\msc{auflia}}

\newcommand{\FOF}{\msc{fof}}

\newcommand{\Zenon}{{Zenon}\xspace}

\newcommand{\CVC}{\msc{CVC4}}
\newcommand{\ZZZ}{\msc{Z3}}

\mdef\id{v}
\mdef\op{w}
\mdef\o{\omega}
\mdef{\xs}{\textbf{x}}
\mdef{\ys}{\textbf{y}}
\mdef\qand{\quad\text{and}\quad}
\mdef\qqand{\qquad\text{and}\qquad}

\newcommand{\lam}[3][x]{\ma{[#1\in#2\mapsto#3]}}

\newcommand{\sep}{\ma{\ |\ }}
\mdef{\lra}{\longrightarrow}
\mdef{\rla}{\longleftarrow}

\newcommand{\trstla}{\mbox{$(\text{\tlaplus},\lra)$}\xspace}

\newcommand{\deq}{\mathrel{\smash   
    {{\stackrel{\scriptscriptstyle\Delta}{=}}}}}
\mdef{\iff}{\Leftrightarrow}
\mdef{\fun}{\rightarrow}
\mdef{\implies}{\Rightarrow}
\newcommand{\tuple}[1]{\ma{\langle#1\rangle}}
\newcommand{\IFTE}[3]{\ma{\IF #1 \THEN #2~\textsc{else}~#3}\xspace}
\newcommand{\IFexp}{\textsc{if-then-else}\xspace}
\newcommand{\Choose}{\mbox{\sc choose}\xspace}
\newcommand{\Union}{\mbox{\sc union}\xspace}

\newcommand{\smt}[1]{\ma{\mathsf{#1}}}
\newcommand{\inttou}{\smt{i2u}}

\makeatletter
\newcommand{\mac}[1]{\relax\ifmmode#1\else{\ensuremath{#1}}\@\expandafter\xspace\fi}
\makeatother
\newcommand{\Int}{\mac{\mathsf{Int}}}
\newcommand{\Bool}{\mac{\mathsf{Bool}}}

\makeatletter
\mdef{\V}{\mathcal{V}}				
\mdef{\O}{\mathcal{O}}				
\mdef{\D}{\mathcal{D}}				

\mdef\U{\mathsf{U}}				
\mdef\A{\forall}
\mdef\E{\exists}
\makeatother

\usepackage{amsthm}
\newtheorem{theorem}{Theorem}

\title{Encoding \Tlaplus set theory\\into many-sorted first-order logic}
\author{Stephan Merz and Hern\'an Vanzetto
  \\[1em] 
  \small{Inria, Villers-l\`es-Nancy, F-54600, France}\\
  \small{Universit\'e de Lorraine, LORIA, UMR 7503, Vandoeuvre-l\`es-Nancy, F-54506, France}\\
  \small{CNRS, LORIA, UMR 7503, Vandoeuvre-l\`es-Nancy, F-54506, France}
}
\date{}

\begin{document}

\maketitle

\begin{abstract}
  We present an encoding of Zermelo-Fraenkel set theory into many-sorted
  first-order logic, the input language of state-of-the-art \SMT solvers.  This
  translation is the main component of a back-end prover based on \SMT solvers
  in the \tlaplus Proof System.
\end{abstract}

\section{Introduction}

The specification language \tlaplus~\cite{lamport:book} combines a variant of
Zermelo-Fraenkel (\ZF) set theory for the description of the data manipulated by
algorithms, and linear-time temporal logic for the specification of their
behavior. The \tlaplus Proof System (\tlaps) provides support for mechanized
reasoning about \tlaplus specifications; it integrates backends for making
automatic reasoners available to users of \tlaps. The work reported here is
motivated by the development of an \SMT backend through which users of \tlaps
interact with off-the-shelf \SMT (satisfiability modulo theories) solvers for
non-temporal reasoning in the set theory of \tlaplus.

More specifically, \tlaps is built around a so-called Proof Manager that
interprets the proofs occurring in the \tlaplus module provided by the user,
generates corresponding proof obligations, and passes them to external automated
verifiers, which are the back-end provers of \tlaps.

Previous to this work, three back-end provers with different capabilities were
available: Isabelle/\tlaplus, a faithful encoding of \tlaplus set theory in the
Isabelle proof assistant, which provides automated proof methods based on
first-order reasoning and rewriting; \Zenon~\cite{zenon}, a tableau prover for
first-order logic with equality that includes extensions for reasoning about
sets and functions; and a backend called SimpleArithmetic, now deprecated,
implementing a decision procedure for Presburger
arithmetic.\footnote{ The backends available prior to the work presented here
  also included a generic translation to the input language of \SMT solvers that
  focused on quantifier-free formulas of linear arithmetic.  This \SMT backend
  was occasionally useful because the other backends perform quite poorly on
  obligations involving arithmetic reasoning. However, it covered only a small
  subset of \tlaplus.  }

The Isabelle and \Zenon backends have very limited support for arithmetic
reasoning, while SimpleArithmetic handles only pure arithmetic formulas,
requiring the user to manually decompose the proofs until the corresponding
proof obligations fall within the respective fragments.  Beyond its integration
as a semi-automatic backend, Isabelle/\tlaplus serves as the most trusted
back-end prover. Accordingly, it is also intended for certifying proof scripts
produced by other back-end provers.  When possible, backends are expected to
produce a detailed proof that can be checked by Isabelle/\tlaplus. Currently,
only the \Zenon backend has an option for exporting proofs that can be certified
in this way.

In this paper we describe the foundations of a back-end prover based on \SMT
solvers for non-temporal proof obligations arising in
\TLAPS.\footnote{Non-temporal reasoning is enough for proving safety properties
  and makes up the vast majority of proof steps in liveness proofs.}
When verifying distributed algorithms, proof obligations are usually ``shallow",
but they still require many details to be checked: interactive proofs can become
quite large without powerful automated back-end provers that can cope with a
significant fragment of the language.  \tlaplus heavily relies on modeling data
using sets and functions. Tuples and records, which occur very often in \tlaplus
specifications, are defined as functions. Assertions mixing first-order logic
(\FOL) with sets, functions, and arithmetic expressions arise frequently in
safety proofs of \tlaplus specifications.  Accordingly, we do not aim at proofs
of deep theorems of mathematical set theory but at good automation for
obligations mixing elementary set expressions, functions, records, and (linear)
integer arithmetic, and our main focus is on \SMT solvers, although we have also
used the techniques described here with \FOL provers.  The de-facto standard
input language for \SMT solvers is \SMTLIB~\cite{smtlib}, which is based on
multi-sorted \FOL (\MSFOL~\cite{manzano:2005}).

In Section~\ref{chap:tla2smt} we present the translation from \tlaplus to
\MSFOL. Although some of the encoding techniques that we use can be found in
similar tools for other set-theoretic languages, the particularities of \tlaplus
make the translation non-trivial:
\begin{itemize}
\item Since \tlaplus is untyped, ``silly'' expressions such as $3 \cup \TRUE$
  are legal; they denote some (unspecified) value. \tlaplus does not even
  distinguish between Boolean and non-Boolean expressions, hence Boolean values
  can be stored in data structures just like any other values.
\item Functions, which are defined axiomatically, are total and have a
  domain. This means that a function applied to an element of its domain has the
  expected value but for any other argument, the value of the function
  application is unspecified. Similarly, the behavior of arithmetic operators is
  specified only for integer arguments.
\item \tlaplus is equipped with a deterministic choice operator (Hilbert's
  $\varepsilon$ operator), which has to be soundly encoded.
\end{itemize}

The first item is particularly challenging for our objectives: whereas an
untyped language is very expressive and flexible for writing specifications,
standard \MSFOL reasoners rely on types for good automation.  In order to
support \tlaplus expressions in a many-sorted environment, we use only one sort
to encode all \tlaplus expressions. We therefore call this translation the
``untyped'' encoding of \tlaplus, where type inference of sorted expressions
such as arithmetic is essentially delegated to the solvers.  In the following we
will use the terms \emph{type} and \emph{sort} interchangeably.

Section~\ref{sec:tlalogic} describes the underlying logic of \tlaplus,
Section~\ref{chap:eval} provides experimental results, and
Section~\ref{chap:conc} concludes and gives directions for future work.


\paragraph{Related work} In previous publications~\cite{LPAR-19,avocs:2012}, we
presented primitive encodings of \tlaplus into \SMTLIB, where \CHOOSE
expressions were not fully supported and Boolification was not made explicit in
the translation.  As a preprocessing step, we developed a type system with
dependent and refinement types for \tlaplus~\cite{nfm:2014}: an algorithm takes
a \tlaplus proof obligation and annotates it with types, which are then used to
simplify our encoding~\cite{avocs:2012}.

Some of the encoding techniques presented in Section~\ref{chap:tla2smt} were
already defined before or are simply folklore, but to our knowledge they have
not been combined and studied in this way. Moreover, the idiosyncrasies of
\tlaplus render their applicability non-trivial. For instance, \tlaplus's
axiomatized functions with domains, including tuples and records, are deeply
rooted in the language.

The Rodin tool set supporting Event-B is based on two translations. The
\textsf{SMT~solvers} plugin~\cite{deharbe:2012} directly encodes simple sets
(\emph{i.e.}, excluding set of sets) as polymorphic $\lambda$-expressions, which
are non-standard and are only handled by the parser of the veriT \SMT solver.
The \textsf{ppTrans} plugin~\cite{pptrans} generates different \SMT sorts for
each basic set and every combination of sets (power sets or cartesian products)
found in the proof obligation.  Similarly, Mentre et al.~\cite{mentre:2012} rely
on Why3 as an interface to discharge Atelier-B proof obligations using different
\SMT solvers, with sets having a polymorphic type.

Recently, Delahaye et al.~\cite{Zenon-Modulo} proposed a different approach to
reason about set theory, instead of a direct encoding into \FOL. The theory of
deduction modulo is an extension of predicate calculus that includes rewriting
of terms and of propositions, and which is well suited for proof search in
axiomatic theories, as it turns axioms into rewrite rules. For example, Peano
arithmetic or Zermelo set theory can be encoded without axioms, turning the
proof search based on axioms into computations.  Zenon
Modulo~\cite{Zenon-Modulo,Jacquel:2012} implements deduction modulo within a
first-order theorem prover.

\msc{mptp}~\cite{urban:2003} translates Mizar to \TPTP/\FOF. The Mizar language
provides second-order predicate variables and abstract terms derived from
replacement and comprehension, such as the set
\[
  \mbox{$\{ n - m \text{ where } m,n \text{ is } Integer : n < m\}$}.
\] 
During preprocessing, \msc{mptp} replaces them by fresh symbols, with their
definitions at the top level.  Similar to our abstraction technique (cf.\
Section~\ref{sec:abstraction}), it is comparable to Skolemization.  In contrast
to our intended application, \msc{mptp} is mainly targeted at mathematical
reasoning.


\section{\tlaplus set theory}
\label{sec:tlalogic}

In this section we describe a fragment of the language of proof obligations
generated by the \tlaplus Proof System that is relevant for this paper. This
language is a variant of \FOL with equality, extended in particular by syntax
for set, function and arithmetic expressions, and a construct for a
deterministic choice operator.  For a complete presentation of the \tlaplus
language see~\cite[Sec.\ 16]{lamport:book}.

We assume given two non-empty, infinite, and disjoint collections $\V$ of
\emph{variable} symbols, and \O of \emph{operator}
symbols,\footnote{ \tlaplus operator symbols correspond to the standard function
  and predicate symbols of first-order logic but we reserve the term
  ``function'' for \tlaplus functional values.}
each equipped with its arity.  The only syntactical category in the language is
the \emph{expression}. For presentational purposes we distinguish between terms,
formulas, set objects, etc. An {expression}~$e$ is inductively defined by the
following grammar:
\[
\begin{array}[t]{r@{}ll}
e ::=~ & \id \sep \op(e,\ldots,e) & \text{(terms)}\\
  \sep & \FALSE \sep e \implies e \sep \A \id\colon\,e
  		 \sep e = e \sep e \in e & \text{(formulas)}\\
  \sep & \{\}
	\sep \{e, e\}
	\sep \SUBSET s
	\sep \UNION s
	\\
  \sep & \{\id \in e : e\}
	\sep \{e : \id \in e\}
	& \text{(sets)}\\
  \sep & \CHOOSE x\colon e 
	& \text{(choice)}\\
  \sep & e[e]
	\sep \DOMAIN e
	\sep \lam[\id]{e}{e}
	\sep [e \fun e]
	& \text{(functions)}\\
  \sep & 0
	\sep 1
	\sep 2
	\sep \ldots
	\sep Int
	\sep Nat
	\sep -e
	\sep e + e
	\sep e < e
	\sep e\ ..\ e
	& \text{(arithmetic)}\\
    \sep & \IFTE{e}{e}{e}
	& \text{(conditional)}\\
  \end{array}
\]

A \emph{term} is a variable symbol~\id in~\V or an application of an operator
symbol~\op in~\O to expressions.  \emph{Formulas} are built from $\FALSE$,
implication and universal quantification, and from the binary operators~$=$
and~$\oldin$.  From these formulas, we can define the familiar constant \TRUE,
the unary~$\neg$ and the binary connectives~$\land$, $\lor$, $\iff$, and the
existential quantifier $\E$.  Also, $\A x \in S \colon e$ is defined as
$\A x \colon x \in S \implies e$.  In standard set theory, sets are constructed
from axioms that state their existence. \tlaplus has explicit syntax for set
objects (empty set, pairing, power set, generalized union, and two forms of set
comprehension derived from the standard axiom schema of replacement), whose
semantics is defined axiomatically.  Since \tlaplus is a set theoretic language,
every expression -- including formulas, functions, numbers, etc.\ -- denotes a
set.

Another primitive construct of \tlaplus is Hilbert's choice
operator~$\varepsilon$, written~$\CHOOSE x\colon P(x)$, that denotes an
arbitrary but fixed value~$x$ such that~$P(x)$ is true, provided that such a
value exists. Otherwise the value of $\CHOOSE x\colon P(x)$ is arbitrary. The
semantics of \Choose is expressed by the following axiom schemas. The first one
gives an alternative way of defining quantifiers, and the second one expresses
that \Choose is deterministic.
\begin{align}
  &\big(\E x \colon P(x)\big) ~\iff~ P\big(\CHOOSE x \colon P(x)\big)
    \label{axiom:choose1}\\
  &\big(\A x \colon P(x) \iff Q(x)\big)~\implies~
   \big(\CHOOSE x \colon P(x)\big) = \big(\CHOOSE x \colon Q(x)\big)
    \label{axiom:choose2}
\end{align}
From axiom~\eqref{axiom:choose2} note that if there is no value satisfying some
predicate $P$, \emph{i.e.}, $\A x\colon P(x) \iff \FALSE$ holds, then
\(
  (\CHOOSE x \colon P(x)) = (\CHOOSE x \colon \FALSE).
\) 
Consequently, the expression $\CHOOSE x \colon \FALSE$ and all its equivalent
forms represent a unique value.

Certain \tlaplus values are \emph{functions}. Unlike standard ZF set theory,
\tlaplus functions are not defined as sets of pairs, but \tlaplus provides
primitive syntax associated with functions.  The expression~$f[e]$ denotes the
result of applying function~$f$ to~$e$, $\DOMAIN f$ denotes the domain of~$f$,
and $\lam{S}{e}$ denotes the function~$g$ with domain~$S$ such that
\mbox{$g[x] = e$}, for any $x \in S$. For $x \notin S$, the value of $g[x]$ is
unspecified.  A \tlaplus value $f$ is a function if and only if it satisfies the
predicate $IsAFcn(f)$ defined as $f = \lam{\DOMAIN f}{f[x]}$.  The fundamental
law governing \tlaplus functions is
\begin{equation}
  f = \lam{S}{e} \,\iff\,
          IsAFcn(f) \land
          \DOMAIN f = S \land
          \A x \in S\colon f[x] = e
  \label{axiom:function}
\end{equation}

Natural numbers $0,1,2, \ldots$ are primitive symbols of \tlaplus. Standard
modules of \tlaplus define $Int$ to denote the set of integer numbers, the
operators $+$ and $<$ are interpreted in the standard way when their arguments
are integers, and the interval $a\,..\,b$ is defined as
$\{n \in Int : a \leq n \land n \leq b\}$.


\section{Untyped encoding of \tlaplus into MS-FOL}
\label{chap:tla2smt}

We define a translation from \tlaplus to multi-sorted first-order logic. Given a
\tlaplus proof obligation, the system generates an equi-satisfiable formula
whose proof can be attempted with automatic theorem provers, including \SMT
solvers.

The translation proceeds in two main steps. First, a preprocessing and
optimization phase applies satisfiability-preserving transformations to a given
\tlaplus formula in order to remove expressions that the target solver cannot
handle. The result is an intermediate \emph{basic} \tlaplus formula,
\emph{i.e.}, a \tlaplus expressions that has an obvious counterpart in the
\SMTLIB/\AUFLIA language.  A basic \tlaplus formula is composed only of \tlaplus
terms and formulas, including equality and set membership relations, plus
primitive arithmetic operators and \IFexp expressions.  All expressions having a
truth value are mapped to the sort~\Bool, and we declare a new sort~\U (for
\tlaplus universe) for all non-Boolean expressions, including sets, functions,
and numbers. Thus, we call this the \emph{untyped} encoding.

\subsection{Boolification}
\label{sec:boolify}

Since \tlaplus has no syntactic distinction between Boolean and non-Boolean
expressions, we first need to determine which expressions are used as
propositions.  We adopt the liberal interpretation of \tlaplus Boolean
expressions where any expression with a top-level connective among logical
operators, $=$, and $\oldin$ has a Boolean
value.\footnote{ The standard semantics of \tlaplus offers three alternatives to
  interpret expressions~\cite[Sec.~16.1.3]{lamport:book}. In the liberal
  interpretation, an expression like $42 \implies \{\}$ always has a truth
  value, but it is not specified if that value is true or false. In the
  conservative and moderate interpretations, the value of $42 \implies \{\}$ is
  completely unspecified. Only in the moderate and liberal interpretation, the
  expression $\FALSE \implies \{\}$ has a Boolean value, and that value is
  true. In the liberal interpretation, all the ordinary laws of logic, such as
  commutativity of $\land$, are valid, even for non-Boolean arguments.  }
Moreover, the result of any expression with a top-level logical connective
agrees with the result of the expression obtained by replacing every argument
$e$ of that connective with $(e = \TRUE)$.

For example, consider the expression $\A x \colon (\neg\neg x) = x$, which is
not a theorem.  Indeed, $x$ need not be Boolean, whereas $\neg\neg x$ is
necessarily Boolean, hence we may not conclude that the expression is
valid. However, $\A x \colon (\neg\neg x) \iff x$ is valid because it is
interpreted as
\(
  \A x \colon (\neg\neg (x=\TRUE)) \iff (x=\TRUE).
\)
Observe that the value of $x=\TRUE$ is a Boolean for any $x$, although the value
is unspecified if $x$ is non-Boolean.

In order to identify the expressions used as propositions we use a simple
algorithm that recursively traverses an expression searching for sub-expressions
that should be treated as formulas. Expressions~$e$ that are used as Booleans,
\emph{i.e.}, that could equivalently be replaced by $e = \TRUE$, are marked as
$e^b$, whose definition can be thought of as $e^b \deq e =\TRUE$. This only
applies if $e$ is a term, a function application, or a \CHOOSE expression.  If
an expression which is known to be non-Boolean by its syntax, such as a set or a
function, is attempted to be Boolified, meaning that a formula is expected in
its place, the algorithm aborts with a ``type'' error.  In \SMTLIB we encode
$x^b$ as $\smt{boolify}(x)$, with $\smt{boolify} : \U \fun \Bool$.  The above
examples are translated as
\(
  \A x^\U \colon (\neg\neg \smt{boolify}(x)) = x
  \text{ and }
  \A x^\Bool \colon (\neg\neg x) \iff x
\) 
and their (in)validity becomes evident.

\subsection{Direct embedding}
\label{sec:encoding}

Our encoding maps in an almost verbatim way Boolified \tlaplus expressions to
corresponding formulas in the target language, without changing substantially
the structure of the original formula. The goal is to encode \tlaplus
expressions using essentially first-order logic and uninterpreted functions. For
first-order \tlaplus expressions it suffices to apply a shallow embedding into
the target language.  Nonlogical \tlaplus operators are declared as function or
predicate symbols with \U-sorted arguments.  For example, the operators~$\cup$
and $\oldin$ are encoded in \SMTLIB as the functions
\(
  \smt{union} : \U \times \U \to \U
\) and \(
  \smt{in} : \U \times \U \to \Bool.
\)

The semantics of standard \tlaplus operators are defined axiomatically. The only
primitive set-theoretical operator is~$\oldin$, so the function \smt{in} will
remain unspecified, while we can express in \MSFOL the axiom for $\cup$ as
\begin{equation}
  \A x^\U, S^\U, T^\U.
  ~~\smt{in}(x,\smt{union}(S,T)) \iff \smt{in}(x,S) \lor \smt{in}(x,T)
  \label{union}
\end{equation}
Note that sets are just values in the universe of discourse (represented by the
sort~\U in the sorted translation), and it is therefore possible to represent
sets of sets and to quantify over sets. The construct for set enumeration
$\{e_1,\ldots,e_n\}$, with $n \geq 0$, is an~$n$-ary expression, so we declare
separate uninterpreted functions for the arities that occur in the proof
obligation, together with the corresponding axioms.

In order to reason about the theory of arithmetic, an automated prover requires
type information, either generated internally, or provided explicitly in the
input language.  The axioms that we have presented so far rely on FOL over
uninterpreted function symbols over the single sort~\U. For arithmetic
reasoning, we want to benefit from the prover's native capabilities. We declare
an unspecified, injective function \(\inttou\is \Int \to \U\)
that embeds built-in integers into the sort~\U.  The typical injectivity axiom
\[
  \A m^\Int,n^\Int \colon \smt{i2u}(m) = \smt{i2u}(n) \implies m = n
\]
generates instantiation patterns for every pair of occurrences of
\smt{i2u}. Noting that \smt{i2u} is injective iff it has a partial inverse, we
use instead the axiom
\(
  \A n^\Int \colon \smt{u2i}(\smt{i2u}(n)) = n,
\)
which generates a linear number of $\smt{i2u}(n)$ instances, where the inverse
$\smt{u2i}: \U \fun \Int$ is unspecified.  Integer literals~$k$ are encoded as
$\inttou(k)$.

For example, the formula $3 \in {Int}$ is translated as
$\smt{in(\inttou(3),tla\_Int)}$ and we have to add to the translation the axiom
for ${Int}$:
\begin{equation} \label{rw:int}
  \A x^\U \colon \smt{in}(x,\smt{tla\_Int}) \iff \E n^\Int \colon x = \inttou(n)
\end{equation}
Observe that this axiom introduces two quantifiers to the translation. We can
avoid the universal quantifier by encoding expressions of the form~$x \in {Int}$
directly into \mbox{$\E n^\Int \colon x = \inttou(n)$}, but the provers would
still have to deal with the existential quantifier.

Arithmetic operators over \tlaplus\ values are defined homomorphically over the
image of \inttou by axioms such as
\begin{equation}
  \label{eq:plus}
  \A m^\Int, n^\Int \colon \smt{plus}(\inttou(m),\inttou(n))\ =\ \inttou(m + n)
\end{equation}
where $+$ denotes the built-in addition over integers. For other arithmetic
operators we define analogous axioms.

In all these cases, type inference is, in some sense, delegated to the back-end
prover. The link between built-in operations and their \tlaplus counterparts is
effectively defined only for values in the range of the function~\inttou.  This
approach can be extended to other useful theories that are natively supported,
such as arrays or algebraic datatypes.

\subsection{Preprocessing and optimizations}
\label{sec:preprocess}

The above encoding has two limitations. First, some \tlaplus expressions cannot
be written in first-order logic. Namely, they are $\{x \in S : P\}$,
\mbox{$\{e : x \in S\}$}, $\CHOOSE x \colon P$, and~\lam{S}{e}, where the
predicate $P$ and the expression~$e$, both of which may have~$x$ as free
variable, become second-order variables when quantified. Secondly, the above
encoding does not perform and scale well in practice.  State-of-the-art \SMT
solvers provide \emph{instantiation patterns} to control the potential explosion
in the number of ground terms generated for instantiating quantified variables,
but we have not been able to come up with patterns to attach to the axiom
formulas that would significantly improve the performance, even for simple
theorems.

What we do instead is to perform several transformations to the \tlaplus proof
obligation to obtain an equi-satisfiable formula which can be straightforwardly
passed to the solvers using the above encoding.

\subsubsection{Normalization}
\label{sec:rewriting}

We define a rewriting process that systematically expands definitions of
non-basic operators. Instead of letting the solver find instances of the
background axioms, it applies the ``obvious'' instances of those axioms during
the translation. In most cases, we can eliminate all non-basic operators.  For
instance, the \ZF axiom for the $\Union$ operator yields the rewriting rule
\[
  x \in \UNION S \lra \exists T \in S\colon x \in T.
\]

All defined rewriting rules apply equivalence-preserving transformations.  We
ensure soundness by proving in Isabelle/\tlaplus that all rewriting rules
correspond to theorems of \tlaplus. The theorem corresponding to a rule
$e \lra f$ is \mbox{$\A \xs: e \iff f$} when $e$ and $f$ are Boolean expressions
and $\A \xs: e = f$ otherwise, where $\xs$ denotes all free variables in the
rule. Most of these theorems exist already in the standard library of
Isabelle/\Tlaplus's library.

The standard extensionality axiom for sets is unwieldy because it introduces an
unbounded quantifier, which can be instantiated by any value of sort \U. We
therefore decided not to include it in the default background theory. Instead,
we instantiate equality expressions $x = y$ whenever possible with the
extensionality property corresponding to $x$ or $y$.  In these cases, we say
that we \emph{expand} equality.  For each set expression $T$ we derive rewriting
rules for equations $x = T$ and $T = x$.  For instance, the rule
\[
  x = \{z \in S : P\} 
  	\lra \A z \colon z \in x \iff z \in S \land P
\]
is derived from set extensionality and the \ZF axiom of bounded set
comprehension.

By not including general extensionality, the translation becomes incomplete.
Even if we assume that the automated theorem provers are semantically complete,
it may happen that the translation of a semantically valid \tlaplus formula
becomes invalid when encoded. In these cases, the user will need to explicitly
add the axiom to the \tlaplus proof.

We also include the rule 
\(
  \A z \colon z \in x \iff z \in y \lra x = y
\)
for the \emph{contraction} of set extensionality, which we apply with higher
priority than the expansion rules.  All above rules of the form $\phi \lra \psi$
define a term rewriting system~\cite{baader:1999} noted \trstla, where $\lra$ is
a binary relation over well-formed \tlaplus expressions.
\begin{theorem} 
  \trstla terminates and is confluent.
\end{theorem}
\vspace{-5pt}
\begin{proof}[Proof (idea)] 
  Termination is proved by embedding \trstla into another reduction system that
  is known to terminate, typically~\mbox{$(\mathbb{N},>)$} \cite{baader:1999}.
  The embedding is through an ad-hoc monotone mapping $\mu$ such that
  $\mu(a) > \mu(b)$ for every rule $a \lra b$. It is defined in such a way that
  every rule instance strictly decreases the number of non-basic and complex
  expressions such as quantifiers or arithmetic expressions.  For confluence, by
  Newman's lemma~\cite{baader:1999}, it suffices to prove that all critical
  pairs are joinable. Thus, we just need to find the critical pairs
  $\tuple{e_1,e_2}$ between all combinations of rewriting rules, and then prove
  that~$e_1$ and~$e_2$ are joinable for each such pair.  In particular, the
  contraction rule is necessary to obtain a strong normalizing system.
\end{proof}

\subsubsection{Functions}
\label{sec:functions}

A \tlaplus function~$\lam{S}{e(x)}$ is akin to a ``bounded''
$\lambda$-abstraction: the function application $\lam{S}{e(x)}[y]$ reduces to
the expected value $e(y)$ if the argument $y$ is an element of $S$, as stated by
the axiom~\eqref{axiom:function}. As a consequence, e.g., the formula
\begin{equation*}
  f = \lam{\{1,2,3\}}{x*x} ~\implies~ f[0] < f[0] + 1,
  \label{example:domaincond}
\end{equation*}
although syntactically well-formed, should not be provable. Indeed, since $0$ is
not in the domain of $f$, we cannot even deduce that $f[0]$ is an integer.

We represent the application of an expression $f$ to another expression~$x$ by
two distinct first-order terms depending on whether the \emph{domain condition}
$x \in \DOMAIN f$ holds or not: we introduce binary operators \a and \o with
conditional definitions
\[
  x \in \DOMAIN f \implies \a(f,x) = f[x]
\qand
  x \notin \DOMAIN f \implies \o(f,x) = f[x].
\]
From these definitions, we can derive the theorem
\begin{equation}
  f[x] = \IFTE{x \in \DOMAIN f}{\a(f,x)}{\o(f,x)}
  \label{fcnapp:2}
\end{equation}
that gives a new defining equation for function application. In this way,
functions are just expressions that are conditionally related to their argument
by~\a and~\o.

The expression $f[0]$ in the above example is encoded as 
\[
  \IFTE{0 \in \DOMAIN f}{\a(f,0)}{\o(f,0)}.
\] 
The solver would have to use the hypothesis to deduce that
$\DOMAIN f = \{1,2,3\}$, reducing the condition $0 \in \DOMAIN f$ to false. The
conclusion can then be simplified to
\mbox{\(
  \o(f,0) < \o(f,0) + 1, 
\)}
which cannot be proved, as expected.  Another example is $f[x] = f[y]$ in a
context where~\mbox{$x = y$} holds: the formula is valid irrespective of whether
the domain conditions hold or not.

Whenever possible, we try to avoid the encoding of function application as in
the definition~\eqref{fcnapp:2}.  From~\eqref{axiom:function}
and~\eqref{fcnapp:2}, we deduce the rewriting rule:
\begin{align}
  \lam{S}{e}[a]
    &\lra 
    \IFTE{a \in S}{e[x \leftarrow a]}{\o(\lam{S}{e},a)}\!
  \label{fcnapp:3}
\end{align}
where $e[x \leftarrow a]$ denotes $e$ with $a$ substituted for $x$.  These rules
replace two non-basic operators (function application and the function
expression) in the left-hand side by only one non-basic operator in the
right-hand side (the first argument of $\o$).

The expression \lam{S}{e} cannot be mapped directly to a first-order
expression. Even in sorted languages like \MSFOL, functions have no notion of
function domain other than the types of their arguments.  Explicit functions
will be treated by the abstraction method below.  What we can do for the moment
is to expand equalities involving functions. The following rewriting rule
derived from axiom~\eqref{axiom:function} replaces the function construct by a
formula containing only basic operators:
\begin{equation*}
  f = \lam{S}{e} \lra
    IsAFcn(f) \land
    \DOMAIN f = S \land
    \A x \in S\colon \a(f,x) = e
\end{equation*}
Observe that we have simplified $f[x]$ by $\a(f,x)$, because $x \in \DOMAIN f$.
This mechanism summarizes the essence of the abstraction method to deal with
non-basic operators described in the next subsection.

In order to prove that two functions are equal, we need to add a background
axiom that expresses the extensionality property for functions:
\[
  \A f,g\colon
  \begin{array}[t]{@{}c@{~}l}
	\land& IsAFcn(f) \land IsAFcn(g)\\
	\land& \DOMAIN f = \DOMAIN g\\
    \land& \A x \in \DOMAIN g \colon \a(f,x)=\a(g,x)\\
    \implies& f = g
  \end{array}
\]
Again, note that $f[x]$ and $g[x]$ were simplified using $\a$. Unlike set
extensionality, this formula is guarded by $IsAFcn$, avoiding the instantiation
of expressions that are not considered functions.  To prove
that~$\DOMAIN f = \DOMAIN g$, we still need to add to the translation the set
extensionality axiom, which we abstain from. Instead, reasoning about the
equality of domains can be solved with an instance of set extensionality for
\DOMAIN expressions only.

\tlaplus defines $n$-tuples as functions with domain $1..n$ and records as
functions whose domain is a finite set of strings. By treating them as non-basic
expressions, we just need to add suitable rewriting rules to \trstla, in
particular those for extensionality expansion.

\subsubsection{Abstraction}
\label{sec:abstraction}

Applying rewriting rules does not always suffice for obtaining formulas in basic
normal form. As a toy example, consider the valid proof obligation
\(
  \A x \colon P(\{x\} \cup \{x\}) \iff P(\{x\}).
\)
The impediment is that the non-basic sub-expressions $\{x\} \cup \{x\}$ and
$\{x\}$ do not occur in the form expected by the left-hand sides of rewriting
rules. They must first be transformed into a form suitable for rewriting.

We call the technique described here \emph{abstraction} of non-basic
expressions.  After applying rewriting, some non-basic expression~$\psi$ may
remain in the proof obligation.  For every occurrence of~$\psi$, we introduce in
its place a fresh term~$y$, and add the formula \mbox{$y = \psi$} as an
assumption in the appropriate context. The new term acts as an
\emph{abbreviation} for the non-basic expression, and the equality acts as its
\emph{definition}, paving the way for a transformation to a basic expression
using the above rewriting rules. Non-basic expressions occurring more than once
are replaced by the same fresh symbol.

In our example the expressions $\{x\} \cup \{x\}$ and~$\{x\}$ are replaced by
fresh constant symbols $k_1(x)$ and $k_2(x)$. Then, the abstracted formula is
\[
  \begin{array}[t]{@{}c@{~}l@{}}
  \land & \A y_1 \colon k_1(y_1) = \{y_1\} \cup \{y_1\}\\
  \land & \A y_2 \colon k_2(y_2) = \{y_2\}\\
  \implies & \A x \colon P(k_1(x)) \iff P(k_2(x)).
  \end{array}
\]
which is now in a form where it is possible to apply the instances of
extensionality to the equalities in the newly introduced definitions.  In order
to preserve satisfiability of the proof obligation, we have to add as hypotheses
instances of extensionality contraction for every pair of definitions where
extensionality expansion was applied. The final equi-satisfiable formula in
basic normal form is
\[
  \begin{array}[t]{@{}c@{~}l@{}}
  \land &\A z,y\colon z \in k_1(y) \iff z = y \lor z = y\\
  \land &\A z,y\colon z \in k_2(y) \iff z = y\\
  \land &\A y_1,y_2 \colon (\A z\colon z \in k_1(y_1) 
    \iff z \in k_2(y_2)) \implies k_1(y_1) = k_2(y_2)\\
  \implies &\A x \colon P(k_1(x)) \iff P(k_2(x)).
  \end{array}
\]

\subsubsection{Eliminating definitions}
\label{sec:simplification}

To improve the encoding, we introduce a procedure that eliminates definitions,
having the opposite effect of the abstraction method where definitions are
introduced and afterwards expanded to basic expressions.  This process collects
definitions of the form~$x = \psi$, and then simply substitutes every occurrence
of the term~$x$ by the non-basic expression~$\psi$ in the rest of the context,
by applying the equality oriented as the rewriting rule~\mbox{$x \lra \psi$}.
The definitions we want to eliminate typically occur in the original proof
obligation, meaning that they are not artificially introduced.  In the following
subsection, we will explain the interplay between normalization, definition
abstraction, and definition elimination.

This transformation produces expressions that can eventually be normalized to
their basic form.  The restriction that $x$ does not occur in $\psi$ avoids
rewriting loops and ensures termination of this process. For instance, the two
equations~\mbox{$x=y$} and~$y=x+1$ will be transformed into~$y=y+1$, which
cannot further be
rewritten.\footnote{The problem of efficiently eliminating definitions from
  propositional formulas is a major open question in the field of proof
  complexity. The definition-elimination procedure can result in an exponential
  increase in the size of the formula when applied
  na\"ively~\cite{avigad:definitions}.}
After applying the substitution, we can safely discard from the resulting
formula the definition~$x = \psi$, when~$x$ is a variable. However, we must keep
the definition if $x$ is an applied operator. Suppose we discard an assumption
$\DOMAIN f = S$, where the conclusion is~$f \in [S \fun T]$. Only after applying
the rewriting rules, the conclusion will be expanded to an expression
containing~$\DOMAIN f$, but the discarded fact required to simplify it to $S$
will be missing.

\subsubsection{Preprocessing algorithm}

\newcommand{\Fix}{\text{\textsc{Fix}}~}

Now we can put together the encoding techniques described above in a single
algorithm that we call \emph{Preprocess}.
\begin{center}
\begin{minipage}{.9\textwidth}
\begin{array}[t]{l@{~}l}
  \emph{Preprocess}(\phi) \defeq
  	\begin{array}[t]{@{}l}
	\phi\\
	\triangleright\ \emph{Boolify}\\
	\triangleright\ \Fix \emph{Reduce}
	\end{array}
  &
  \emph{Reduce}(\phi) \defeq
	\begin{array}[t]{@{}l} 
      \phi\\
	  \triangleright\ \Fix (\emph{Eliminate} \circ \emph{Rewrite})\\
	  \triangleright\ \Fix (\emph{Abstract} \circ \emph{Rewrite})\\
	\end{array}  
\end{array}
\end{minipage}
\end{center}
Here, $\textsc{Fix}\ \mathcal{A}$ means that step $\mathcal{A}$ is executed
until reaching a fixed point, the combinator $\triangleright$, used to chain
actions on a formula $\phi$, is defined as $\phi \triangleright f \deq f(\phi)$,
and function composition $\circ$ is defined as
$f \circ g \deq \lambda\phi.\,g(f(\phi))$.

The \textit{Preprocess} algorithm takes a \tlaplus formula $\phi$, Boolifies it
and then applies repeatedly the step called \emph{Reduce}, until reaching a
fixed point, to transform the formula into a basic normal form. Only then the
resulting formula is ready to be translated to the target language using the
embedding of Section~\ref{sec:encoding}. In turn, \textit{Reduce} first
eliminates the definitions in the given formula
(Sect.~\ref{sec:simplification}), applies the rewriting rules
(Sect.~\ref{sec:rewriting}) repeatedly, and then applies abstraction
(Sect.~\ref{sec:abstraction}) followed by rewriting repeatedly. Observe that the
elimination step is in some sense opposite to the abstraction step: the first
one eliminates every definition~$x = \psi$ by using it as the rewriting
rule~$x \lra \psi$, while the latter introduces a new symbol~$x$ in the place of
an expression~$\psi$ and asserts~$x = \psi$, where $\psi$ is non-basic in both
cases. Therefore, elimination should only be applied before abstraction, and
each of those should be followed by rewriting.

The \textit{Preprocess} algorithm is sound because it is composed of sound
sub-steps. It also terminates, meaning that it will always compute a basic
normal formula, but with a caveat: we have to be careful that \textit{Abstract}
and \textit{Eliminate} do not repeatedly act on the same
expression. \textit{Eliminate} does not produce non-basic expressions, but
\textit{Abstract} generates definitions that can be processed by
\textit{Eliminate}, reducing them again to the original non-basic
expression. That is the reason for \textit{Rewrite} to be applied after every
application of \textit{Abstract}: the new definitions are rewritten, usually by
an extensionality expansion rule. In short, termination depends on the existence
of extensionality rewriting rules for each kind of non-basic expression that
\textit{Abstract} may catch. Then, for any \tlaplus expression there exists an
equi-satisfiable basic expression in normal form that the algorithm will
compute.

\subsection{Encoding \Choose}
\label{sec:choose}

The \Choose operator is notoriously difficult for automatic provers to reason
about. Nevertheless, we can exploit \Choose expressions by using the axioms that
define them.  By introducing a definition for $\CHOOSE x\colon P(x)$, we obtain
the theorem
\[
  \big(y = \CHOOSE x \colon P(x)\big) \implies 
  	\big((\E x \colon P(x)) \iff P(y)\big),
\]
where $y$ is some fresh symbol. This theorem can be conveniently used as a
rewriting rule after abstraction of \Choose expressions, and for \Choose
expressions that occur negatively, in particular, as hypotheses of proof
obligations.

For determinism of choice (axiom~\eqref{axiom:choose2}), suppose an arbitrary
pair of \Choose expressions
\[
  \p1 \deq \CHOOSE x \colon P(x)
\qand
  \p2 \deq \CHOOSE x \colon Q(x)
\]
where the free variables of \p1 are $x_1,\ldots,x_n$ (noted~\xs) and those of
\p2 are $y_1,\ldots,y_m$ (noted~\ys).  We need to check whether formulas~$P$
and~$Q$ are equivalent for every pair of expressions~\p1 and~\p2 occurring in a
proof obligation. By abstraction of \p1 and \p2, we obtain the axiomatic
definitions
\(
  \A \xs \colon f_1(\xs) = \CHOOSE x \colon P(x)
\) and \(
  \A \ys \colon f_2(\ys) = \CHOOSE x \colon Q(x),
\)
where $f_1$ and $f_2$ are fresh operator symbols of suitable arity.  Then, we
state the extensionality property for the pair $f_1$ and $f_2$ as the axiom
\[
  \A \xs,\ys \colon
  \big(\A x \colon P(x) \iff Q(x)\big) \implies f_1(\xs) = f_2(\ys).
\]


\section{Evaluation}
\label{chap:eval}

In order to validate our approach we reproved several test cases that had been
proved interactively using the previously available \tlaps back-end provers,
namely \Zenon, Isabelle/\tlaplus and the decision procedure for Presburger
arithmetic. We will refer to the combination of those three backends as ZIP for
short.

\begin{table}[t]
\centering

\begin{tabular}{l@{\;}|@{\;}r@{\;}|r*{2}{|r@{\;}r@{\;}}}
\multicolumn{2}{l|}{} & 
    \multicolumn{1}{c}{ZIP} & 
    \multicolumn{2}{|c}{CVC4} & 
    \multicolumn{2}{|c}{Z3}
  \\
\cline{1-2}
\cline{4-7}
\multicolumn{1}{c}{} 
& \multicolumn{1}{@{}c@{}|}{size}
& 
& u & t
& u & t
\\
\hline
{Peterson} &
3   & -
	      &  0.41 &   0.46
	      &  {0.34} &   0.40
\\
{Peterson} &
10  & 5.69
	      &  {0.78} &  0.96
	      &  0.80 &  0.97
\\
\hline
{Bakery} &
 16 & -
	    & - &    6.57
	    & - &  7.15
\\
{Bakery} &
 223 & 52.74
 & &
\\
\hline
{Memoir}-T & 1 & -
  &     - &     - 
  &  1.99 &  1.53
\\
{Memoir}-T & 12 & -
  & 3.11 &  3.46
  & 3.21 &  3.51
\\
{Memoir}-T & 424 & 7.31
  &&
  &&
\\\hline
{Memoir}-I & 8 & -
  & 3.84 & 5.79
  & 9.35 & 10.23
\\
{Memoir}-I & 61 & 8.20
  & & 
\\\hline
{Memoir}-A & 27 & -
  & 11.31 & 14.36
  & 11.46 & 14.30
\\
{Memoir}-A & 126 & 19.10
  & & 
\end{tabular}
\medskip

\begin{tabular}{l|r|r|r|rr}
\multicolumn{1}{c|}{Finite sets} & 
    \multicolumn{2}{c}{ZIP} & 
    \multicolumn{3}{|c}{Zenon+SMT}
  \\
\cline{1-1}
\cline{2-6}
\multicolumn{1}{c}{} 
& size & 
& size
& u & t
\\
\hline
CardZero & 11 & 5.42
  & 5 
  & 0.48 & 0.48 
\\
CardPlusOne & 39 & 5.35
  & 3
  & 0.49 & 0.52
\\
CardOne & 6 & 5.36
  & 1
  & 0.35 & 0.35
\\
CardOneConv & 9 & 0.63
  & 2 & 0.35 & 0.36
\\
FiniteSubset & 62 & 7.16 
 & 19 & - & 5.77
\\
PigeonHole & 42 & 7.07
 & 20 & 7.01 & 7.22
\\
CardMinusOne & 11 & 5.44 
  & 5 & 0.75 & 0.73
\end{tabular}\\
\caption{Evaluation benchmarks results. An entry with the symbol ``-" means 
  that the solver has reached the timeout without finding the proof for at 
  least one of the proof obligations. The backends were executed with a 
  timeout of 300 seconds.}
\label{tbl:results}
\end{table}

For each benchmark, we compare two dimensions of an interactive proof: size and
time. We define the \emph{size} of an interactive proof as the number of
non-trivial proof obligations generated by the Proof Manager. This number is
proportional to the number of interactive steps and therefore represents the
user effort for making \tlaps check the proof. The \emph{time} is the number of
seconds required by the Proof Manager to verify those proofs on a standard
laptop.

Table~\ref{tbl:results} presents the results for four case studies: mutual
exclusion proofs of the Peterson and Bakery algorithms, type-correctness and
refinement proofs of the Memoir security architecture~\cite{memoir:report}, and
proofs of theorems about the cardinality of finite sets.  We compare how proofs
of different sizes are handled by the backends.  Each line corresponds to an
interactive proof of a given size. Columns correspond to the running times for a
given \SMT solver, where each prover is executed on all generated proof
obligations. For our tests we have used the state-of-the-art \SMT solvers \CVC
v1.3 and \ZZZ v4.3.0. For each prover we present two different times
corresponding to the untyped encoding (the column labeled~\smt{u}) and the
optimized encoding using the type system with refinement types~\cite{nfm:2014}
(labeled~\smt{t}).

In all cases, the use of the new backend leads to significant reductions in
proof sizes and running times compared to the original interactive proofs. In
particular, the ``shallow'' proofs of the first three case studies required only
minimal interaction. We have also used the new \SMT backend with good success on
several proofs not shown here.  Both \SMT solvers offer similar results, with
\ZZZ being better at reasoning about arithmetic. In a few cases \CVC is faster
or even proves obligations on which \ZZZ fails.  Some proof obligations can be
proved only by \Zenon, in the case of big structural high-level formulas, or
only using the ``typed'' encoding, because heavy arithmetic reasoning is
required.


\section{Conclusions}
\label{chap:conc}

We have presented a sound and effective way of discharging \tlaplus proof
obligations using automated theorem provers based on many-sorted first-order
logic. This encoding was implemented in a back-end prover that integrates
external \SMT solvers as oracles to the \tlaplus Proof System \tlaps.  The main
component of the backend is a generic translation framework that makes available
to \tlaps any \SMT solver that supports the de facto standard format
\SMTLIB/\AUFLIA. We have also used the same framework for integrating automated
theorem provers based on unsorted \FOL, such as those based on the superposition
calculus.

The resulting translation can handle a useful fragment of the \tlaplus language,
including set theory, functions, linear arithmetic expressions, and the \CHOOSE
operator (Hilbert's choice). Encouraging results show that \SMT solvers
significantly reduce the effort of interactive reasoning for verifying
``shallow'' \tlaplus proof obligations, as well as some more involved formulas
including linear arithmetic expressions. Both the size of the interactive proof,
which reflects the number of user interactions, and the time required to find
automatic proofs can be remarkably reduced with the new back-end prover.

The mechanism that combines term-rewriting with abstraction enables the backend
to successfully handle \CHOOSE expressions, tuples, records, and \tlaplus
functions ($\lambda$-abstractions with domains).  However, our rewriting method
may introduce many additional quantifiers, which can be difficult for the
automated provers to handle.

The untyped universe of \tlaplus is represented as a universal sort in
\MSFOL. Purely set-theoretic expressions are mapped to formulas over
uninterpreted symbols, together with relevant background axioms. The built-in
integer sort and arithmetic operators are homomorphically embedded into the
universal sort, and type inference is in essence delegated to the solver.  The
soundness of the encoding is immediate: all the axioms about sets, functions,
records, tuples, etc.\ are theorems in the background theory of \tlaplus that
exist in the Isabelle encoding. The ``lifting'' axioms for the encoding of
arithmetic assert that \tlaplus arithmetic coincides with \SMT arithmetic over
integers. For ensuring completeness of our encoding, we would have to include
the standard axiom of set extensionality in the background theory. For
efficiency reasons, we include only instances of extensionality for specific
sets, function domains, and functions.

The translation presented here forms the basis for further
optimizations. In~\cite{nfm:2014} we have explored the use of (incomplete) type
synthesis for \tlaplus expressions, based on a type system with dependent and
refinement types. Extensions for reasoning about real arithmetic and finite
sequences would be useful. More importantly, we rely on the soundness of
external provers, temporarily including them as part of \tlaps's trusted
base. In future work we intend to reconstruct within Isabelle/\tlaplus (along
the lines presented in~\cite{sledgehammer-smt}) the proof objects that many \SMT
solvers can produce. Such a reconstruction would have to take into account not
only the proofs generated by the solvers, but also all the steps performed
during the translation, including rewriting and abstraction.

\bibliographystyle{abbrv}
\bibliography{all}

\end{document}